\documentclass[conference,a4paper]{IEEEtran}

\usepackage[utf8]{inputenc} 
\usepackage[T1]{fontenc}
\usepackage{url}              
\usepackage{cite}             

\usepackage[cmex10]{amsmath}  
\usepackage{mleftright}       
\mleftright                   

\usepackage{graphicx}         
\usepackage{booktabs}         

\usepackage{tikz}
\usepackage{theorem}
\usepackage{amssymb}
\usepackage{amsfonts, dsfont}
\usepackage{enumitem}
\allowdisplaybreaks

\usepackage{tikz-cd}
\usetikzlibrary{calc,arrows,positioning}
\tikzstyle myBG=[line width=3pt,opacity=1]

\DeclareMathOperator*{\supp}{supp}

\DeclareMathOperator*{\Unif}{Unif}

\DeclareMathOperator*{\interior}{int}

\newtheorem{lemma}{Lemma}
\newtheorem{theorem}{Theorem}[section]
\newtheorem{definition}[theorem]{Definition}
\newtheorem{corollary}[theorem]{Corollary}
\newtheorem{proposition}[theorem]{Proposition}

\newtheorem{remark}[theorem]{Remark}

\def\proof{\noindent{\it Proof}. \ignorespaces}
\def\endproof{\vbox{\hrule height0.6pt\hbox{\vrule height1.3ex%
width0.6pt\hskip0.8ex\vrule width0.6pt}\hrule height0.6pt}}

\def\ar#1{\textcolor{red}{\footnotesize {(ar: #1)}}}

\begin{document}

\title{Complementary Graph Entropy, AND Product, and Disjoint Union of Graphs} 



\author{\IEEEauthorblockN{Nicolas Charpenay}
\IEEEauthorblockA{\textit{Institut de Recherche en Informatique}\\
\textit{et Systèmes Aléatoires (IRISA)}\\
Rennes, FRANCE \\
nicolas.charpenay@irisa.fr}
\and
\IEEEauthorblockN{Ma\"el le Treust}
\IEEEauthorblockA{\textit{Institut de Recherche en Informatique}\\
\textit{et Systèmes Aléatoires (IRISA)} \\
Rennes, FRANCE \\
mael.le-treust@irisa.fr}
\and
\IEEEauthorblockN{Aline Roumy}
\IEEEauthorblockA{\textit{Institut National de Recherche} \\
\textit{en Informatique et en Automatique (INRIA)}\\
Rennes, FRANCE \\
aline.roumy@inria.fr}
}

%
%

%
%

\maketitle

\begin{abstract}
In the zero-error Slepian-Wolf source coding problem, the optimal rate is given by the complementary graph entropy $\overline{H}$ of the characteristic graph. It has no single-letter formula, except for perfect graphs, for the pentagon graph with uniform distribution $G_5$, and for their disjoint union. We consider two particular instances, where the characteristic graphs respectively write as an AND product $\wedge$, and as a disjoint union $\sqcup$. We derive a  structural result that equates $\overline{H}(\wedge \: \cdot)$ and $\overline{H}(\sqcup \: \cdot)$ up to a multiplicative constant, which has two consequences. First, we prove that the cases where $\overline{H}(\wedge \:\cdot)$ and $\overline{H}(\sqcup \: \cdot)$ can be linearized coincide. 
Second, we determine $\overline{H}$ in cases where it was unknown:
products of perfect graphs; and $G_5 \wedge G$ when $G$ is a perfect graph, using Tuncel et al.'s result for $\overline{H}(G_5 \sqcup G)$. The graphs in these cases are not perfect in general.
\end{abstract}

\section{Introduction}\label{section:intro}

We study the zero-error variant of Slepian and Wolf source coding problem depicted in Figure \ref{fig:SWzero}, where the estimate $\widehat{X}^n$ must be equal to $X^n$ with probability one. This problem is also called ``restricted inputs'' in Alon and Orlitsky's work \cite{alon1996source}. 

\subsection{Characteristic graphs and optimal rate $\overline{H}$}

\begin{figure}[t!]
    \centering
    \begin{tikzpicture}
    \node[shape=rectangle, draw=black, fill=white, inner sep=5pt] (E) at (0,0) {Encoder};
    \node[shape=rectangle, draw=black, fill=white, inner sep=5pt] (D) at (3,0) {Decoder};

    \node[draw=none] (SY) at ($(D)+(0,-.75)$) {$Y^n$};
    \node[draw=none] (X_1) at ($(D)+(2,0)$) {$\widehat{X}^n = X^n$};
    \node[draw=none] (X) at ($(E)+(-1.5,0)$) {$X^n$};
    
    \draw[->, >=stealth] (D) edge (X_1);
    \draw[->, >=stealth] (X) edge (E);
    \draw[->, >=stealth] (SY) edge (D);
    \draw[->, >=stealth] (E) edge (D);
    
    \node[draw=none] (R0) at ($(1.5,0)$) {$\diagup$};
    \node[draw=none] (R) at ($(R0)+(0,10pt)$) {$R$};
    
    \end{tikzpicture}
    \caption{Zero-error Slepian-Wolf source coding problem.}
    \label{fig:SWzero}
\end{figure}
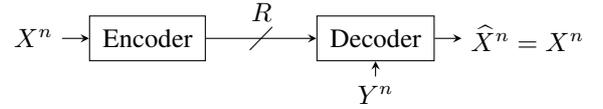

An adequate probabilistic graph $G$ (i.e. a graph with an underlying probability distribution on its vertices) can be associated to a given instance of zero-error source coding problem in Figure \ref{fig:SWzero}, as in Witsenhausen's work \cite{witsenhausen1976zero}. This graph is called ``characteristic graph'' of the problem, as it encompasses the problem data in its structure: the vertices are the source alphabet, with the source probability distribution $P_X$ on these vertices, and two source symbols $xx'$ are adjacent if they are ``confusable'', i.e. $P_{X,Y}(x,y)P_{X,Y}(x',y) > 0$ for some side information symbol $y$. By construction, the encoder must map adjacent symbols in $G$ to different codewords in order to prevent any decoding error: the colorings of the graph $G$ directly correspond to zero-error encoding mappings.

The best rate that can be achieved in the problem of Figure \ref{fig:SWzero} with $n = 1$ is the minimal entropy of the colorings of $G$, as shown in \cite{alon1996source}. This quantity is called \emph{chromatic entropy} and is denoted by 
\begin{align}
   H_{\chi}(G)\doteq \inf \lbrace H(c(V)) \:|\: c \text{ is a coloring of } G \rbrace.
\end{align}

The asymptotic optimal rate in the problem of Figure \ref{fig:SWzero} is characterized by
\begin{align}
        \overline{H}(G) = \lim_{n \rightarrow \infty} \frac{1}{n} H_{\chi}(G^{\wedge n}),\label{eq:debitopti}
\end{align}
where $G^{\wedge n}$ is the $n$-iterated \emph{AND product} of the characteristic graph $G$, see \cite{alon1996source}.
As shown in \cite{koulgi2003zero}, it is equal to the \emph{complementary graph entropy} defined in \cite{korner1973two}.

A single-letter formula for $\overline{H}$ is not known, except for perfect graphs \cite{csiszar1990entropy}; and for $G_5 \sqcup G$ and its complement, for all perfect graph $G$ \cite{tuncel2009complementary}, where $G_5$ is the pentagon graph with uniform distribution. 

\subsection{Characteristic graph structure in particular instances}

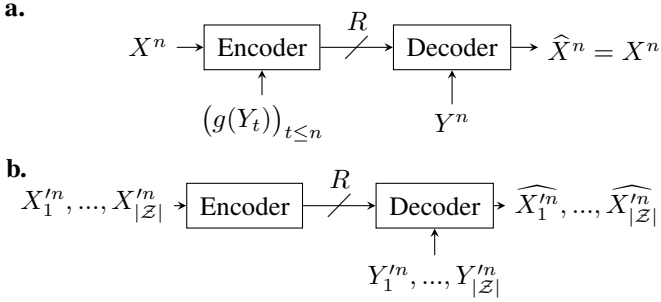
\begin{figure}[t!]
    \centering
    \begin{tikzpicture}
    \node[shape=rectangle, draw=black, fill=white, inner sep=5pt] (E) at (0,0) {Encoder};
    \node[shape=rectangle, draw=black, fill=white, inner sep=5pt] (D) at (2.5,0) {Decoder};

    \node[draw=none] (SY) at ($(D)+(0,-1)$) {$Y^n$};
    \node[draw=none] (X_1) at ($(D)+(2,0)$) {$\widehat{X}^n = X^n$};
    \node[draw=none] (G) at ($(E)+(0,-1)$) {$\big(g(Y_t)\big)_{t \leq n}$};
    \node[draw=none] (X) at ($(E)+(-1.5,0)$) {$X^n$};
    
    \draw[->, >=stealth] (D) edge (X_1);
    \draw[->, >=stealth] (X) edge (E);
    \draw[->, >=stealth] (G) edge (E);
    \draw[->, >=stealth] (SY) edge (D);
    \draw[->, >=stealth] (E) edge (D);
    
    \node[draw=none] (R0) at ($(1.25,0)$) {$\diagup$};
    \node[draw=none] (R) at ($(R0)+(0,10pt)$) {$R$};
    
    \node[draw=none] (A) at ($(E)+(-3.25,.5)$) {\textbf{a.}};
    
    \end{tikzpicture}
    \begin{tikzpicture}
    \node[shape=rectangle, draw=black, fill=white, inner sep=5pt] (E) at (0,0) {Encoder};
    \node[draw=none] (B) at ($(E)+(-3,.5)$) {\textbf{b.}};
    \node[shape=rectangle, draw=black, fill=white, inner sep=5pt] (D) at (2.5,0) {Decoder};
    
    \node[draw=none] (SY) at ($(D)+(0,-1)$) {$Y'^n_1, ..., {Y'^n_{|\mathcal{Z}|}}$};
    \node[draw=none] (X_1) at ($(D)+(2,0)$) {$\widehat{X'^n_1}, ..., \widehat{X'^n_{|\mathcal{Z}|}}$};
    \node[draw=none] (X) at ($(E)+(-2,0)$) {$X'^n_1, ..., {X'^n_{|\mathcal{Z}|}}$};
    
    \draw[->, >=stealth] (D) edge (X_1);
    \draw[->, >=stealth] (X) edge (E);
    \draw[->, >=stealth] (SY) edge (D);
    \draw[->, >=stealth] (E) edge (D);
    
    \node[draw=none] (R0) at ($(1.25,0)$) {$\diagup$};
    \node[draw=none] (R) at ($(R0)+(0,10pt)$) {$R$};
    
    \end{tikzpicture}
    \caption{Two particular instances of zero-error Slepian-Wolf source coding problem, where $g: \mathcal{Y} \rightarrow \mathcal{Z}$ is deterministic, $(X'^n_z, Y'^n_z) \sim P^n_{X,Y|g(Y) = z}$ for all $z \in \mathcal{Z}$, and the pairs $((X'^n_z,Y'^n_z))_{z \in \mathcal{Z}}$ are mutually independent. For all $z \in \mathcal{Z}$, the auxiliary graph $G_{z}$ is Witsenhausen's characteristic graph for the pair $(X'_z, Y'_z)$.}
    \label{fig:Setting}
\end{figure}

Since determining $\overline{H}$ is difficult, let us consider particular instances of the problem in Figure \ref{fig:SWzero}, depicted in Figure \ref{fig:Setting}. Both settings have a characteristic graph with a specific structure. Thanks to the side information at the encoder in Figure \ref{fig:Setting}.a, the characteristic graph is the \emph{disjoint union} ($\sqcup$) of a family of auxiliary probabilistic graphs $(G_{z})_{z \in \mathcal{Z}}$; and in Figure \ref{fig:Setting}.b the characteristic graph is the \emph{AND product} ($\wedge$) of the $(G_{z})_{z \in \mathcal{Z}}$. Both $\sqcup$ and $\wedge$ are binary operators on probabilistic graphs that play a central role in this study. A natural question arises in the context of Figure \ref{fig:Setting}: can we determine the optimal rates if we only know $\overline{H}(G_{z})$ for all $z \in \mathcal{Z}$? With the subadditivity results in \cite[Theorem 2]{tuncel2009complementary}, we know that $\overline{H}\big(\bigsqcup_{z \in \mathcal{Z}}^{P_{g(Y)}} G_{z}\big) \leq \sum_{z \in \mathcal{Z}} P_{g(Y)} \overline{H}(G_{z})$ and $\overline{H}\big(\bigwedge_{z \in \mathcal{Z}} G_{z}\big) \leq \sum_{z \in \mathcal{Z}} \overline{H}(G_{z})$ holds in general, 
however characterizing the cases where equality holds is an open problem.

\subsection{Related work}

If the decoder wants to recover a function $f(X,Y)$ instead of $X$, the setting of Figure \ref{fig:SWzero} becomes the zero-error variant of the ``coding for computing'' problem \cite{orlitsky1995coding}. Charpenay et al. study in \cite{charpenay2022zero} the variant with side information at the encoder, i.e. the setting from Figure \ref{fig:Setting}.a with $f(X,Y)$ requested by the decoder. In \cite{ravi2015zero}, Ravi and Dey study a setting with a bidirectional relay. In \cite{malak2022fractional}, Malak introduces a fractional version of chromatic entropy in a lossless coding for computing scenario.

Another important problem is the Shannon capacity $\Theta$ of a graph \cite{shannon1956zero}, which characterizes the optimal rate in the zero-error channel coding scenario. Marton has shown in \cite{marton1993shannon} that $\overline{H}(G) + C(G,P) = H(P)$, where $P$ is the underlying probability distribution of $G$, and $C(G,P)$ is the graph capacity relative to $P$. 
The same questions on linearization arise for $\Theta$: for which $G,G'$ do we have $\Theta(G \wedge G') = \Theta(G) \Theta(G')$? A counterexample is shown by Haemers in \cite{haemers1979some}, using an upper-bound on $\Theta$ based on the rank of the adjacency matrix. Refinements of Haemers bound are developed in \cite{bukh2018fractional} by Bukh and Cox, and in \cite{gao2022tracial} by Gao et al. Recently in \cite{schrijver2023shannon}, Schrijver shows that $\Theta(G \wedge G') = \Theta(G) \Theta(G')$ is equivalent to $\Theta(G \sqcup G') = \Theta(G) + \Theta(G')$. The computability of $\Theta$ is investigated in \cite{boche2021computability} by Boche and Deppe. An asymptotic expression for $\Theta$ using semiring homomorphisms is given by Zuiddam et al. in \cite{zuiddam2018algebraic}. In \cite{gu2021non}, Gu and Shayevitz study the two-way channel case. An extension of $\Theta$ for secure communication is developed in \cite{wiese2018secure} by Wiese et al. 

\subsection{Contributions}

In this paper we link the complementary graph entropies of a disjoint union of probabilistic graphs with that of their product, i.e. $\overline{H}(\sqcup \:\cdot\:)$ and $\overline{H}(\wedge \:\cdot\:)$. 
First, we show a structural result on the complementary graph entropy of a disjoint union w.r.t. a type $P_A$, that makes use of $\wedge$ instead of $\sqcup$. This enables us to equate $\overline{H}(\sqcup \:\cdot\:)$ and $\overline{H}(\wedge \:\cdot\:)$ up to a multiplicative constant. This formula has several consequences.

Firstly, we can derive with it a single-letter formula $\overline{H}$ of products of perfect graphs. This case was unsolved as a product of perfect graphs is not perfect in general. However, a disjoint union of perfect graphs is perfect, this is why studying disjoint unions is the key. 
Finally, it enables us to show that the linearizations of $\overline{H}(\sqcup \:\cdot\:)$ and $\overline{H}(\wedge \:\cdot\:)$ are equivalent; i.e. if equality holds for either equation in Tuncel et al.'s subadditivity results \cite[Theorem 2]{tuncel2009complementary}, then equality also holds for the other one. We use this result to determine the complementary graph entropy of the non-perfect probabilistic graph $G_5 \wedge G$ when $G$ is perfect.


In Section \ref{section:defs}, we define the graph-theoretic concepts we need to formulate our main theorems in Section \ref{sec:main}, and their consequences in Section \ref{section:consequences}. An example of application for these theorems is given in Section \ref{section:example}, and the main proofs are developed in Section \ref{section:prooflemmapartvalues}, Section \ref{section:proofmainths} and Section \ref{section:proofthmainperfect}.

\section{Notations and definitions}\label{section:defs}

We denote sequences by $x^n = (x_1, ..., x_n)$. 

The set of probability distributions over $\mathcal{X}$ is denoted by $\Delta(\mathcal{X})$; $P_X \in \Delta(\mathcal{X})$ is the distribution of a random variable $X$. The uniform distribution is denoted by $\text{Unif}$. The conditional distribution of $X$ knowing $Y$ is denoted by $P_{X|Y}$.

A probabilistic graph $G$ is a tuple $(\mathcal{V}, \mathcal{E}, P_V)$, where $(\mathcal{V}, \mathcal{E})$ is a graph and $P_V \in \Delta(\mathcal{V})$.
A subset $\mathcal{S} \subseteq \mathcal{V}$ is independent in $G$ if for all $x,x' \in \mathcal{S}$, $xx' \notin \mathcal{E}$. A mapping $c : \mathcal{V} \rightarrow \mathcal{C}$ is a coloring if $c^{-1}(i)$ is independent for all $i \in \mathcal{C}$. The cycle, complete, and empty graphs with $n$ vertices are respectively denoted by $C_n$, $K_n$, $N_n$. 

\begin{definition}[AND product $\wedge$]
The AND product of $G_1 = (\mathcal{V}_1, \mathcal{E}_1, P_{V_1})$ and $G_2 = (\mathcal{V}_2, \mathcal{E}_2, P_{V_2})$ is a probabilistic graph denoted by $G_1 \wedge G_2$ with: 
\begin{itemize}[label = -]
    \item $\mathcal{V}_1 \times \mathcal{V}_2$ as set of vertices,
    \item $P_{V_1}P_{V_2}$ as probability distribution on the vertices,
    \item $(v_1 v_2),(v'_1 v'_2)$ are adjacent if $v_1v'_1 \in \mathcal{E}_1$ AND $v_2v'_2 \in \mathcal{E}_2$; with the convention of self-adjacency for all vertices.
\end{itemize}
We denote by $G_1^{\wedge n}$ the $n$-th AND power: $G_1^{\wedge n} \doteq G_1 \wedge ... \wedge G_1$.\vspace{-.7em}
\end{definition} 

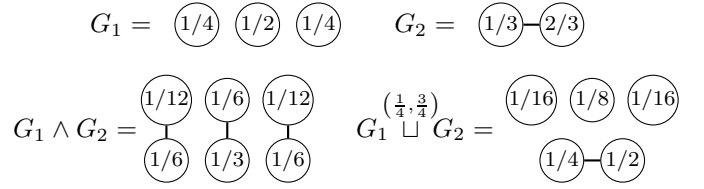
\begin{figure}[t!]
    \centering
    \begin{tikzpicture}[scale=.8]
        \node[shape=circle,draw=black, inner sep = .5pt,fill=white] (0) at ($(0,0)$) {\footnotesize$1/4$};
        \node[shape=circle,draw=black, inner sep = .5pt,fill=white] (1) at ($(0) + (1,0)$) {\footnotesize$1/2$};
        \node[shape=circle,draw=black, inner sep = .5pt,fill=white] (2) at ($(0) + (2,0)$) {\footnotesize$1/4$};
        \node[draw=none] (N) at ($(0)+(-1.25,0)$) {$G_1 = $};
        
        \node[shape=circle,draw=black, inner sep = .5pt,fill=white] (3) at ($(5,0)$) {\footnotesize$1/3$};
        \node[shape=circle,draw=black, inner sep = .5pt,fill=white] (4) at ($(3)+(1,0)$) {\footnotesize$2/3$};
        \node[draw=none] (K) at ($(3)+(-1.25,0)$) {$G_2 = $};
        \draw[draw=black,thick] (4) edge (3);
        
        \node[shape=circle,draw=black, inner sep = .5pt,fill=white] (04) at ($(-.5,-2.25)$) {\footnotesize$1/6$};
        \node[shape=circle,draw=black, inner sep = 0pt,fill=white] (03) at ($(04)+(0,1)$) {\footnotesize$1/12$};
        \node[shape=circle,draw=black, inner sep = .5pt,fill=white] (14) at ($(04)+(1,0)$) {\footnotesize$1/3$};
        \node[shape=circle,draw=black, inner sep = .5pt,fill=white] (13) at ($(04)+(1,1)$) {\footnotesize$1/6$};
        \node[shape=circle,draw=black, inner sep = .5pt,fill=white] (24) at ($(04)+(2,0)$) {\footnotesize$1/6$};
        \node[shape=circle,draw=black, inner sep = 0pt,fill=white] (23) at ($(04)+(2,1)$) {\footnotesize$1/12$};
        \node[draw=none] (W) at ($(04)+(-1.5,0.5)$) {$G_1 \wedge G_2 = $};
        \draw[draw=black,thick] (04) edge (03);
        \draw[draw=black,thick] (14) edge (13);
        \draw[draw=black,thick] (24) edge (23);
        
        \node[shape=circle,draw=black, inner sep = 0pt,fill=white] (0_) at ($(5.5,-1.25)$) {\footnotesize$1/16$};
        \node[shape=circle,draw=black, inner sep = .5pt,fill=white] (1_) at ($(0_) + (1,0)$) {\footnotesize$1/8$};
        \node[shape=circle,draw=black, inner sep = 0pt,fill=white] (2_) at ($(0_) + (2,0)$) {\footnotesize$1/16$};
        \node[shape=circle,draw=black, inner sep = .5pt,fill=white] (3_) at ($(0_) + (.5,-1)$) {\footnotesize$1/4$};
        \node[shape=circle,draw=black, inner sep = .5pt,fill=white] (4_) at ($(0_)+(1.5,-1)$) {\footnotesize$1/2$};
        \node[draw=none] (U) at ($(0_)+(-1.75,-.25)$) {$G_1\!\!\! \overset{\left(\frac{1}{4},\frac{3}{4}\right)}{\sqcup} \!\!\! G_2 = $};
        \draw[draw=black,thick] (3_) edge (4_);

    \end{tikzpicture}
    \caption{An empty graph $G_1 = (N_3, (\frac{1}{4},\frac{1}{2}, \frac{1}{4}) )$ and a complete graph $G_2 = (K_2, (\frac{1}{3}, \frac{2}{3}) )$, along with their AND product $G_1 \wedge G_2$ and their disjoint union $G_1 \sqcup G_2$ w.r.t. $(\frac{1}{4}, \frac{3}{4})$.}
    \label{fig:exprod}
\end{figure}

\begin{definition}[Disjoint union $\sqcup$ of probabilistic graphs] \label{def:disjointunion}
Let $\mathcal{A}$ be a finite set, and let $P_A \in \Delta(\mathcal{A})$. For all $a \in \mathcal{A}$, let $G_a = (\mathcal{V}_a, \mathcal{E}_a, P_{V_a})$ be a probabilistic graph, their disjoint union w.r.t. $P_A$ is a probabilistic graph $(\mathcal{V}, \mathcal{E}, P_V)$ denoted by $\bigsqcup^{P_A}_{a \in \mathcal{A}} G_a$ and defined by:
\begin{itemize}[label = - ]
    \item $\mathcal{V} = \bigsqcup_{a \in \mathcal{A}} \mathcal{V}_a$ is the disjoint union of the sets $(\mathcal{V}_a)_{a \in \mathcal{A}}$;
    \item For all $v,v' \in \mathcal{V}$, $vv' \in \mathcal{E}$ iff they both belong to the same $\mathcal{V}_a$ and $vv' \in \mathcal{E}_a$;
    \item $P_V = \sum_{a \in \mathcal{A}} P_A(a) P_{V_a}$; note that the $(P_{V_a})_{a \in \mathcal{A}}$ have disjoint support in $\mathcal{V}$.
\end{itemize}
\end{definition}

\begin{remark}
    The disjoint union $\sqcup$ that we consider here is also called ``sum of graphs'' by Tuncel et al. in \cite{tuncel2009complementary}. Note that $\sqcup$ is the disjoint union over the vertices: it differs in nature from the union over the edges $\cup$ that is already studied in the literature, in particular in \cite{korner1988graphs}, \cite{csiszar1990entropy} and \cite{marton1993shannon}. 
\end{remark}

An example of AND product and disjoint union is given in Figure \ref{fig:exprod}.

\section{Main result}\label{sec:main}

In this section, $\mathcal{A}$ is a finite set, $P_A$ is a distribution from $\Delta(\mathcal{A})$ and $(G_a)_{a \in \mathcal{A}}$ is a family of probabilistic graphs. 


In Theorem \ref{th:partvalues} we give an expression for the complementary graph entropy of a disjoint union w.r.t. a type; the proof is given in Section \ref{section:proofthpartvalues}. With Corollary \ref{cor:unif} we equate $\overline{H}(\sqcup \:\cdot\:)$ and $\overline{H}(\wedge \:\cdot\:)$ up to a multiplicative constant when $P_A = \Unif(\mathcal{A})$.

\begin{definition}[Type of a sequence]
    Let $a^k \in \mathcal{A}^k$, its type $T_{a^k}$ is its empirical distribution. The set of types of sequences from $\mathcal{A}^k$ is denoted by $\Delta_k(\mathcal{A}) \subset \Delta(\mathcal{A})$.
\end{definition}

\begin{theorem}\label{th:partvalues}
    If $P_A \in \Delta_k(\mathcal{A})$ for some $k \in \mathbb{N}^\star$ then
    \begin{equation}
        \overline{H}\left(\bigsqcup_{a \in \mathcal{A}}^{P_A} G_a\right) = \frac{1}{k}\overline{H}\left(\bigwedge_{a \in \mathcal{A}} G_a^{\wedge k P_A(a)}\right).\label{eq:corpartvalues}
    \end{equation}
\end{theorem}

\begin{corollary}\label{cor:unif}
$\overline{H}\big(\bigsqcup_{a \in \mathcal{A}}^{\Unif(\mathcal{A})} G_a\big) = \frac{1}{|\mathcal{A}|}\overline{H}\left(\bigwedge_{a \in \mathcal{A}} G_a\right)$.
\end{corollary}

\subsection{Proof of Theorem \ref{th:partvalues}}\label{section:proofthpartvalues}

In order to complete the proof, we need Lemma \ref{lemma:partvalues}, it is the cornerstone of the connection between $\overline{H}(\sqcup \:\cdot\:)$ and $\overline{H}(\wedge \:\cdot\:)$. The main reasons why $\wedge$ appears in \eqref{eq:lemmapartvaluesA} are the AND powers used in $\overline{H}$, and the distributivity of $\wedge$ w.r.t. $\sqcup$ (see Lemma \ref{lemma:distrib}). The proof of Lemma \ref{lemma:partvalues} is developed in Section \ref{section:prooflemmapartvalues}.

\begin{lemma}\label{lemma:partvalues}
    Let $(\overline{a}_n)_{n \in \mathbb{N}^\star} \in \mathcal{A}^{\mathbb{N}^\star}$ be any sequence such that $T_{\overline{a}^n} \rightarrow P_A$ when $n \rightarrow \infty$. Then we have
    \begin{equation}
        \overline{H}\left(\bigsqcup_{a \in \mathcal{A}}^{P_A} G_a \right) = \lim_{n \rightarrow \infty} \frac{1}{n}H_\chi\left(\bigwedge_{a \in \mathcal{A}} G_a^{\wedge n T_{\overline{a}^n} (a)}\right).\label{eq:lemmapartvaluesA}
    \end{equation}
\end{lemma}

Now let us prove Theorem \ref{th:partvalues}. Let $(\overline{a}_n)_{n \in \mathbb{N}^\star}$ be a $k$-periodic sequence such that $T_{\overline{a}^k} = P_A$, then $T_{\overline{a}^{nk}} = T_{\overline{a}^k}$ for all $n \in \mathbb{N}^\star$, and $T_{\overline{a}^n} \underset{n \rightarrow \infty}{\rightarrow} P_A$. We can use Lemma \ref{lemma:partvalues} and consider every $k$-th term in the limit:
\begin{align}
    \overline{H}\Big(\textstyle\bigsqcup_{a \in \mathcal{A}}^{P_A} G_a\Big) & = \lim\limits_{n \rightarrow \infty} \frac{1}{kn}H_\chi\Big(\textstyle\bigwedge_{a \in \mathcal{A}} G_a^{\wedge kn T_{\overline{a}^{kn}} (a)}\Big) \nonumber\\
    & = \lim\limits_{n \rightarrow \infty} \frac{1}{kn}H_\chi\Big(\Big(\textstyle\bigwedge_{a \in \mathcal{A}} G_a^{\wedge k T_{\overline{a}^{k}} (a)}\Big)^{\wedge n}\Big) \nonumber\\
    & = \frac{1}{k} \overline{H}\Big(\textstyle\bigwedge_{a \in \mathcal{A}} G_a^{\wedge k P_A (a)}\Big). &\endproof \nonumber
\end{align}

\section{Consequences}\label{section:consequences}

\subsection{Single-letter formula of $\overline{H}$ for products of perfect graphs}\label{section:mainperfect}

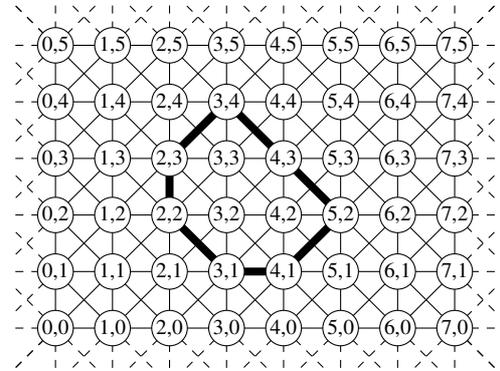
\begin{figure}[b!]
    \centering
    \begin{tikzpicture}[scale=.75]
        \foreach \i in {0,1,2,3,4,5,6}{
        \foreach \j in {0,1,2,3,4}{
        \draw[] ($(\i,\j)$) -- ($(\i+1,\j+1)$);
        \draw[] ($(\i+1,\j)$) -- ($(\i,\j+1)$);
        \draw[] ($(\i,\j)$) -- ($(\i+1,\j)$);
        \draw[] ($(\i,\j)$) -- ($(\i,\j+1)$);
        }
        }
        \foreach \i in {0,1,2,3,4,5,6}{
        \draw[] ($(\i,5)$) -- ($(\i+1,5)$);
        }
        \foreach \j in {0,1,2,3,4}{
        \draw[] ($(7,\j)$) -- ($(7,\j+1)$);
        }
        
        \draw[myBG] ($(2,2)$) -- ($(2,3)$) -- ($(3,4)$) -- ($(4,3)$)  -- ($(5,2)$) -- ($(4,1)$) -- ($(3,1)$) -- ($(2,2)$);
        
        \foreach \i in {0,1,2,3,4,5,6,7}{
        \draw[dashed] ($(\i,5)$) -- ($(\i+.75,5.75)$);
        \draw[dashed] ($(\i,5)$) -- ($(\i,5.75)$);
        \draw[dashed] ($(\i,5)$) -- ($(\i-.75,5.75)$);
        \draw[dashed] ($(\i,0)$) -- ($(\i+.75,-.75)$);
        \draw[dashed] ($(\i,0)$) -- ($(\i,-.75)$);
        \draw[dashed] ($(\i,0)$) -- ($(\i-.75,-.75)$);
        }
        
        \foreach \j in {0,1,2,3,4,5}{
        \draw[dashed] ($(0,\j)$) -- ($(-.75,\j-.75)$);
        \draw[dashed] ($(0,\j)$) -- ($(-.75,\j)$);
        \draw[dashed] ($(0,\j)$) -- ($(-.75,\j+.75)$);
        \draw[dashed] ($(7,\j)$) -- ($(7.75,\j-.75)$);
        \draw[dashed] ($(7,\j)$) -- ($(7.75,\j)$);
        \draw[dashed] ($(7,\j)$) -- ($(7.75,\j+.75)$);
        }
        
        \foreach \i in {0,1,2,3,4,5,6,7}{
        \foreach \j in {0,1,2,3,4,5}{
        \node[shape=circle,draw=black, inner sep = .5pt,fill=white] (\i,\j) at ($(\i,\j)$) {\footnotesize\i,\j};
        }
        }
        
    \end{tikzpicture}
    \caption{This is the AND product of two perfect graphs $C_6$ and $C_8$. The thick edges represent an induced subgraph $C_7$, which makes $C_6 \wedge C_8$ non perfect by the strong perfect graph Theorem (see Theorem \ref{th:strongperfect}).}
    \label{fig:counterperfect}
\end{figure}

With the exceptions of $G_5 = (C_5, \Unif(\lbrace 1, ..., 5\rbrace))$ and $G_5 \sqcup G$ and its complement when $\overline{H}(G)$ is known, the only cases where $\overline{H}$ is known are perfect graphs with any underlying distribution: it is given by the Körner graph entropy, defined below. 
We extend the known cases with Theorem \ref{th:mainperfect}, which gives a single-letter expression for $\overline{H}$ for AND products of perfect graphs. This case was not solved before, as a product of perfect graphs is not perfect in general (see Figure \ref{fig:counterperfect} for a counterexample). The proof of Theorem \ref{th:mainperfect} is developed in Section \ref{section:proofthmainperfect}. 

\begin{definition}[Induced subgraph]
    The subgraph induced in a graph $G$ by a subset of vertices $\mathcal{S}$ is the graph obtained from $G$ by keeping only the vertices in $\mathcal{S}$ and the edges between them, and is denoted by $G[\mathcal{S}]$. When $G$ is a probabilistic graph, we give it the underlying probability distribution $P_V/P_V(\mathcal{S})$.
\end{definition}

\begin{definition}[Perfect graph]
    A graph $G = (\mathcal{V}, \mathcal{E})$ is perfect if $\:\forall \mathcal{S} \subset \mathcal{V}, \; \chi(G[\mathcal{S}]) = \omega(G[\mathcal{S}])$; where $\omega$ is the size of the largest clique (i.e. complete induced subgraph); and $\chi(G[\mathcal{S}])$ is the smallest $|\mathcal{C}|$ such that there exists a coloring $c : \mathcal{S} \rightarrow \mathcal{C}$ of $G[\mathcal{S}]$. By extension, we call perfect a probabilistic graph $(\mathcal{V}, \mathcal{E}, P_{V})$ if $(\mathcal{V}, \mathcal{E})$ is perfect.
\end{definition}

\begin{definition}[Körner graph entropy $H_\kappa$]
For all $G = (\mathcal{V}, \mathcal{E}, P_V)$, let $\Gamma(G)$ be the collection of independent sets of vertices in $G$. The K\"{o}rner graph entropy of $G$ is defined by
\begin{align}
    H_\kappa(G) = \min_{V \in W \in \Gamma(G)} I(W;V), \label{eq:wheee}
\end{align}
where the minimum is taken over all distributions $P_{W|V} \in \Delta(\mathcal{W})^{\mathcal{V}}$, with $\mathcal{W} = \Gamma(G)$ and with the constraint that the random vertex $V$ belongs to the random independent set $W$ with probability one, i.e. $V \in W \in \Gamma(G)$ in \eqref{eq:wheee}.
\end{definition}

\begin{theorem}[Strong perfect graph theorem, from \cite{chudnovsky2006strong}]\label{th:strongperfect}
A graph $G$ is perfect if and only if neither $G$ nor its complement have an induced odd cycle of length at least 5.
\end{theorem}

\begin{theorem}[from \cite{csiszar1990entropy}]\label{th:Hbarperfect}
    Let $G$ be a perfect probabilistic graph, then $\overline{H}(G) = H_\kappa(G)$.
\end{theorem}

\begin{theorem}\label{th:mainperfect}
    When $(G_a)_{a \in \mathcal{A}}$ is a family of perfect probabilistic graphs, the following single-letter characterizations hold:
    \begin{align}
        & \overline{H}\left(\bigwedge_{a \in \mathcal{A}} G_a \right) = \sum_{a \in \mathcal{A}} \overline{H}(G_a) = \sum_{a \in \mathcal{A}} H_\kappa(G_a), \\
        &  \overline{H}\left(\bigsqcup_{a \in \mathcal{A}}^{P_A} G_a \right) = \sum_{a \in \mathcal{A}} P_A(a) \overline{H}(G_a) = \sum_{a \in \mathcal{A}} P_A(a) H_\kappa(G_a).
    \end{align}
\end{theorem}

\subsection{Linearization of the complementary graph entropy}

In their subadditivity result \cite[Theorem 2]{tuncel2009complementary}, Tuncel et al. show that for all probabilistic graphs $G_1,G_2$ and $\alpha \in (0,1)$,
\begin{align}
    & \overline{H}(G_1 \overset{(\alpha, 1-\alpha)}{\sqcup} G_2) \leq \alpha \overline{H}(G_1) + (1-\alpha)\overline{H}(G_2), \label{eq:tuncelA}\\
    & \overline{H}(G_1 \wedge G_2) \leq \overline{H}(G_1) + \overline{H}(G_2).\label{eq:tuncelB}
   \end{align}
We show in Theorem \ref{th:caraclinspec} that the cases where equality holds in \eqref{eq:tuncelA} and \eqref{eq:tuncelB} coincide. 

\begin{theorem}\label{th:caraclinspec}
    For all probabilistic graphs $G_1, G_2$, for all $\alpha \in (0,1)$, we have:
    \begin{align}
        & \overline{H}(G_1 \overset{(\alpha, 1-\alpha)}{\sqcup} G_2) = \alpha \overline{H}(G_1) + (1-\alpha)\overline{H}(G_2) \\
        \Longleftrightarrow\: & \overline{H}(G_1 \wedge G_2) = \overline{H}(G_1) + \overline{H}(G_2).
    \end{align}
\end{theorem}

We prove and use the more general formula stated in Theorem \ref{th:caraclin}. The proof is given in Section \ref{section:proofmainths}.

\begin{theorem}\label{th:caraclin}
    Let $P_A \in \Delta(\mathcal{A})$ with full-support, then the following equivalence holds
    \begin{align}
        & \overline{H}\left(\bigsqcup_{a \in \mathcal{A}}^{P_A} G_a\right) = \sum_{a \in \mathcal{A}} P_A(a) \overline{H}(G_a) \label{eq:cormain01a} \\
        \Longleftrightarrow \;\; & \overline{H}\left(\bigwedge_{a \in \mathcal{A}} G_a\right) = \sum_{a \in \mathcal{A}} \overline{H}(G_a). \label{eq:cormain01b}
    \end{align}
\end{theorem}

A case where equality holds in \eqref{eq:cormain01a} is developed by Tuncel et al. in \cite[Lemma 3]{tuncel2009complementary}: $G_5 \doteq (C_5, \Unif(\lbrace 1, ..., 5\rbrace))$ along with any perfect graph. We provide a single-letter formula for $\overline{H}(G_5 \wedge G)$ when $G$ is perfect; while $G_5 \wedge G$ is not perfect as $G_5 \wedge G$ contains an induced $C_5$ (see Theorem \ref{th:strongperfect}). The proof of the following Corollary is given in Appendix \ref{section:proofcorC_5}.

\begin{corollary}\label{cor:C_5}
    For all perfect probabilistic graph $G$, 
    \begin{align}
        \overline{H}\big(G \wedge G_5) = \overline{H}(G) + \overline{H}(G_5) = H_\kappa(G) + \textstyle\frac{1}{2}\log 5.
    \end{align}
\end{corollary}

\section{Example}\label{section:example}

In this section, for all $i \in \mathbb{N}^\star$, $G_i$ denotes the cycle graph with $i$ vertices uniform distribution, i.e. $G_i = \big(C_i, \Unif(\lbrace 0, ..., i-1\rbrace)\big)$. Both $G_6$ and $G_8$ are perfect, and as shown in Figure \ref{fig:counterperfect}, $G_6 \wedge G_8$ is not a perfect graph. We have:
\begin{align}
    H_\kappa(G_6) & = H(V_6) - \max_{V_6 \in W_6 \in \Gamma(G_6)} H(V_6|W_6)\label{eq:exampleA1} \\
    & = 1 + \log 3 - \log 3 = 1
\end{align}
as $H(V_6|W_6)$ in
\eqref{eq:exampleA1} is maximized by taking $W_6 = \lbrace 0,2,4 \rbrace$ when $V_6 \in \lbrace 0,2,4 \rbrace$, and $W_6 = \lbrace 1,3,5 \rbrace$ otherwise.

Similarly, $H_\kappa(G_8) = 1$.

We can use Theorem \ref{th:Hbarperfect} to find $\overline{H}(G_6\wedge G_8)$:
\begin{align}
    \overline{H}(G_6\wedge G_8) = H_\kappa(G_6) + H_\kappa(G_8) = 2.
\end{align}

We can build an optimal coloring of $G_6 \wedge G_8$, $c^* : (v_6,v_8) \mapsto (\mathds{1}_{v_6 \text{ is even}}, \mathds{1}_{v_8 \text{ is even}})$.

\section{Proof of Lemma \ref{lemma:partvalues}}\label{section:prooflemmapartvalues}

\subsection{Preliminary results}

Lemma \ref{lemma:distrib} establishes the distributivity of $\wedge$ w.r.t. $\sqcup$ for probabilistic graphs, similarly as in \cite{zuiddam2018algebraic} for graphs without underlying distribution. 
Lemma \ref{lemma:simplificationinduced} states that $\overline{H}$ can be computed with subgraphs induced by sets that have an asymptotic probability one, in particular we will use it with typical sets of vertices. The proofs of Lemma \ref{lemma:distrib} and Lemma \ref{lemma:simplificationinduced} are respectively given in Appendix \ref{section:prooflemmadistrib} and Appendix \ref{section:prooflemmasimplificationinduced}.

\begin{lemma}\label{lemma:distrib}
    Let $\mathcal{A}, \mathcal{B}$ be finite sets, let $P_A \in \Delta(\mathcal{A})$ and $P_B \in \Delta(\mathcal{B})$. For all $a \in \mathcal{A}$ and $b \in \mathcal{B}$, let $G_a = (\mathcal{V}_a, \mathcal{E}_a, P_{V_a})$ and $G_b = (\mathcal{V}_b, \mathcal{E}_b, P_{V_b})$ be probabilistic graphs. Then 
    \begin{align}
        \left(\bigsqcup_{a \in \mathcal{A}}^{P_A} G_a\right) \wedge \left(\bigsqcup_{b \in \mathcal{B}}^{P_B} G_b\right) = \bigsqcup_{(a,b) \in \mathcal{A} \times \mathcal{B}}^{P_AP_B} G_a \wedge G_b.\label{eq:distribunion}
    \end{align}
\end{lemma}

\begin{lemma}\label{lemma:simplificationinduced}
    Let $G = (\mathcal{V}, \mathcal{E}, P_V)$, and $(\mathcal{S}^n)_{n \in \mathbb{N}^\star}$ be a sequence of sets such that for all $n \in \mathbb{N}^\star$, $\mathcal{S}^n \subset \mathcal{V}^n$, and $P^n_V(\mathcal{S}^n) \rightarrow 1$ when $n \rightarrow \infty$. Then $\overline{H}(G) = \lim_{n \rightarrow \infty} \frac{1}{n} H_\chi\big(G^{\wedge n}[\mathcal{S}^n]\big)$.
\end{lemma}

\begin{definition}[Isomorphic probabilistic graphs]
Let $G_1 = (\mathcal{V}_1, \mathcal{E}_1, P_{V_1})$ and $G_2 = (\mathcal{V}_2, \mathcal{E}_2, P_{V_2})$. We say that $G_1$ is isomorphic to $G_2$ if there exists an isomorphism between them, i.e. a bijection $\psi : \mathcal{V}_1 \rightarrow \mathcal{V}_2$ such that:
\begin{itemize}[label = - ]
    \item For all $v_1, v_1' \in \mathcal{V}_1$, $v_1 v'_1 \in \mathcal{E}_1 \Longleftrightarrow \psi(v_1)\psi(v'_1) \in \mathcal{E}_2$,
    \item For all $v_1 \in \mathcal{V}_1$, $P_{V_1}(v_1) = P_{V_2}\big(\psi(v_1)\big)$.
\end{itemize}
\end{definition}

\begin{lemma}[from \cite{charpenay2022zero}]\label{lemma:unionisomorph}
    Let $\mathcal{B}$ be a finite set, let $P_B \in \Delta(\mathcal{B})$ and let $(G_b)_{b \in \mathcal{B}}$ be a family of isomorphic probabilistic graphs, then $H_\chi\big(\bigsqcup_{b' \in \mathcal{B}}^{P_B} G_{b'} \big) = H_\chi(G_b)$ for all $b \in \mathcal{B}$.
\end{lemma}

\subsection{Main proof of Lemma \ref{lemma:partvalues}}

For all $a \in \mathcal{A}$, let $G_a = (\mathcal{V}_a, \mathcal{E}_a, P_{V_A})$, and let $G = \bigsqcup^{P_A}_{a \in \mathcal{A}} G_a$. Let $P_A \in \Delta(\mathcal{A})$, and let $(\overline{a}_n)_{n \in \mathbb{N}^\star} \in \mathcal{A}^{\mathbb{N}^\star}$ be a sequence such that $T_{\overline{a}^n} \rightarrow P_A$ when $n \rightarrow \infty$.

Let $\epsilon > 0$, and for all $n \in \mathbb{N}^\star$ let 
\begin{align}
    & \mathcal{T}^n_\epsilon(P_A) \doteq \big\lbrace a^n \in \mathcal{A}^n \:\big|\: \| T_{a^n} - P_A \|_\infty \leq \epsilon \big\rbrace, \label{eq:lemmapartA0} \\
    & P'^n \doteq \frac{P^n_A}{P^n_A(\mathcal{T}^n_\epsilon(P_A))},\qquad \mathcal{S}_\epsilon^n \doteq \bigsqcup_{a^n \in \mathcal{T}^n_\epsilon(P_A)} \;\prod_{t \leq n} \mathcal{V}_{a_t}. \nonumber
\end{align}
By Lemma \ref{lemma:simplificationinduced} we have 
\begin{align}
      \overline{H}(G) = \lim_{n \rightarrow \infty} \frac{1}{n}H_\chi\Big(G^{\wedge n}[\mathcal{S}^n_\epsilon]\Big),\label{eq:conclemmapart1}
\end{align}
as $P^n_V(\mathcal{S}^n_\epsilon) \rightarrow 1$ when $n \rightarrow \infty$. Let us study the limit in \eqref{eq:conclemmapart1}. For all $n$ large enough, $\overline{a}^n \in \mathcal{T}^n_\epsilon(P_A)$ as $T_{\overline{a}^n} \rightarrow P_A$. Therefore, for all $a^n \in \mathcal{T}^n_\epsilon(P_A)$ and $a' \in \mathcal{A}$,
\begin{align}
    \big|T_{\overline{a}^n}(a')-T_{a^n}(a')\big|\leq 2\epsilon.\label{eq:lemmapartB0}
\end{align}

We have on one hand
\begin{align}
    & H_\chi\Big(\big(\textstyle\bigsqcup^{P_A}_{a \in \mathcal{A}} G_a\big)^{\wedge n}[\mathcal{S}^n_\epsilon]\Big) \nonumber\\
    =  \; & H_\chi\left(\left(\textstyle\bigsqcup_{a^n \in \mathcal{A}^n}^{P_A^n} \; \textstyle\bigwedge_{t \leq n} G_{a_t}\right)[\mathcal{S}^n_\epsilon]\right) \label{eq:lemmapartB1} \\
    = \; & H_\chi\left(\textstyle\bigsqcup_{a^n \in \mathcal{T}^n_\epsilon(P_A)}^{P'^n} \;\textstyle\bigwedge_{t \leq n} G_{a_t}\right) \label{eq:lemmapartB2} \\
    = \; & H_\chi\left(\textstyle\bigsqcup_{a^n \in \mathcal{T}^n_\epsilon(P_A)}^{P'^n} \; \textstyle\bigwedge_{a' \in \mathcal{A}} G_{a'}^{\wedge n T_{a^n}(a')}\right) \label{eq:lemmapartB3} \\
    \leq \; & H_\chi\left(\textstyle\bigsqcup_{a^n \in \mathcal{T}^n_\epsilon(P_A)}^{P'^n} \; \textstyle\bigwedge_{a' \in \mathcal{A}} G_{a'}^{\wedge n T_{\overline{a}^n}(a') + \lceil 2 n\epsilon \rceil}\right) \label{eq:lemmapartB4} \\
    = \; & H_\chi\left(\textstyle\bigwedge_{a' \in \mathcal{A}} G_{a'}^{\wedge n T_{\overline{a}^n}(a') + \lceil 2 n\epsilon \rceil}\right) \label{eq:lemmapartB5} \\
    \leq \; & H_\chi\left(\textstyle\bigwedge_{a' \in \mathcal{A}} G_{a'}^{\wedge n T_{\overline{a}^n}(a')}\right) + H_\chi\left(\textstyle\bigwedge_{a' \in \mathcal{A}} G_{a'}^{\wedge \lceil 2 n\epsilon\rceil}\right) \label{eq:lemmapartB5a} \\
    \leq \; & H_\chi\left(\textstyle\bigwedge_{a' \in \mathcal{A}} G_{a'}^{\wedge n T_{\overline{a}^n}(a')}\right) + \lceil 2 n\epsilon \rceil|\mathcal{A}|\log|\mathcal{V}|;\label{eq:lemmapartB6}
\end{align}
where \eqref{eq:lemmapartB1} comes from Lemma \ref{lemma:distrib}; \eqref{eq:lemmapartB2} comes from the definition of $\mathcal{S}_\epsilon^n$ and $P'^n$ in \eqref{eq:lemmapartA0}; \eqref{eq:lemmapartB3} is a rearrangement of the terms inside the product; \eqref{eq:lemmapartB4} comes from \eqref{eq:lemmapartB0}; \eqref{eq:lemmapartB5} follows from Lemma \ref{lemma:unionisomorph}, the graphs $\big(\bigwedge_{a' \in \mathcal{A}} G_{a'}^{\wedge n T_{\overline{a}^n}(a') + \lceil 2 n\epsilon \rceil}\big)_{a^n \in \mathcal{T}^n_\epsilon(P_A)}$ are isomorphic as they do not depend on $a^n$; \eqref{eq:lemmapartB5a} follows from the subadditivity of $H_\chi$; and \eqref{eq:lemmapartB6} is the upper bound on $H_\chi$ given by the highest entropy of a coloring.

On the other hand, we obtain with similar arguments
\begin{align}
    & H_\chi\Big(\big(\textstyle\bigsqcup^{P_A}_{a \in \mathcal{A}} G_a\big)^{\wedge n}[\mathcal{S}^n_\epsilon]\Big) \nonumber\\
    \geq \: & H_\chi\left(\textstyle\bigwedge_{a' \in \mathcal{A}} G_{a'}^{\wedge n T_{\overline{a}^n}(a') - \lceil 2 n\epsilon \rceil}\right)\label{eq:lemmapartC3}\\
    \geq \: & H_\chi\left(\textstyle\bigwedge_{a' \in \mathcal{A}} G_{a'}^{\wedge n T_{\overline{a}^n}(a')}\right) - H_\chi\left(\textstyle\bigwedge_{a' \in \mathcal{A}} G_{a'}^{\wedge \lceil 2 n\epsilon\rceil}\right),\label{eq:lemmapartC4a}\\
    \geq \: & H_\chi\left(\textstyle\bigwedge_{a' \in \mathcal{A}} G_{a'}^{\wedge n T_{\overline{a}^n}(a')}\right) - \lceil 2 n\epsilon \rceil|\mathcal{A}|\log|\mathcal{V}|.\label{eq:lemmapartC5}
\end{align}
Note that \eqref{eq:lemmapartC4a} also comes from the subadditivity of $H_\chi$ : $H_\chi(G_2) \geq H_\chi(G_1 \wedge G_2) - H_\chi(G_1)$ for all $G_1, G_2$.

By combining \eqref{eq:lemmapartB6} and \eqref{eq:lemmapartC5} we obtain
\begin{align}
    & \left|\lim_{n \rightarrow \infty} \frac{1}{n} H_\chi(G^{\wedge n}[\mathcal{S}^n_\epsilon]) - \lim_{n \rightarrow \infty} \frac{1}{n} H_\chi\left(\bigwedge_{a' \in \mathcal{A}} G_{a'}^{\wedge n T_{\overline{a}^n}(a')}\right)\right| \nonumber \\
    & \leq 2\epsilon|\mathcal{A}|\log|\mathcal{V}|.\label{eq:conclemmapart2}
\end{align}

As this holds for all $\epsilon > 0$, combining \eqref{eq:conclemmapart1} and \eqref{eq:conclemmapart2} yields the desired result.

\section{Proof of Theorem \ref{th:caraclin}}\label{section:proofmainths}

\subsection{Preliminary results}

In Lemma \ref{th:etaconvex} we give regularity properties of $P_A \mapsto \overline{H}\big(\bigsqcup_{a \in \mathcal{A}}^{P_A} G_a\big)$. The proof of Lemma \ref{th:etaconvex} is developed in Appendix \ref{section:proofetaconvex}. Lemma \ref{lemma:convexanalysis} states that if a convex function $\gamma$ of $\Delta(\mathcal{A})$ meets the linear interpolation of the $(\gamma(\mathds{1}_a))_{a \in \mathcal{A}}$ at an interior point, then $\gamma$ is linear. We use it for proving the equivalence in Theorem \ref{th:caraclin}, by considering $\gamma = P_A \mapsto \overline{H}\big(\bigsqcup_{a \in \mathcal{A}}^{P_A} G_a\big)$. The proof of Lemma \ref{lemma:convexanalysis} is given in Appendix \ref{section:prooflemmaconvexanalysis}.

\begin{lemma}\label{th:etaconvex}
    The function $P_A \mapsto \overline{H}\big(\bigsqcup_{a \in \mathcal{A}}^{P_A} G_a\big)$ is convex and $(\log \max_a |\mathcal{V}_a|)$-Lipschitz.
\end{lemma}

\begin{lemma}\label{lemma:convexanalysis}
    Let $\mathcal{A}$ be a finite set, and $\gamma : \Delta(\mathcal{A}) \rightarrow \mathbb{R}$ be a convex function. Then the following holds:
    \begin{align}
        & \exists P_A \in \interior(\Delta(\mathcal{A})), \, \gamma(P_A) = \textstyle\sum_{a \in \mathcal{A}} P_A(a) \gamma(\mathds{1}_a) \label{eq:lemma1A}\\
        \Longleftrightarrow \;\; & \forall P_A \in \Delta(\mathcal{A}), \, \gamma(P_A) = \textstyle\sum_{a \in \mathcal{A}} P_A(a) \gamma(\mathds{1}_a) \label{eq:lemma1B}
    \end{align}
    where $\interior(\Delta(\mathcal{A}))$ is the interior of $\Delta(\mathcal{A})$ (i.e. the full-support distributions on $\mathcal{A}$).
\end{lemma}

\subsection{Main proof of Theorem \ref{th:caraclin}}

$(\Longrightarrow)$ Assume that $\overline{H}\left(\textstyle\bigwedge_{a \in \mathcal{A}} G_{a}\right) = \textstyle\sum_{a \in \mathcal{A}} \overline{H}(G_{a})$.

We can use Corollary \ref{cor:unif}: $\overline{H}\big(\bigsqcup_{a \in \mathcal{A}}^{\Unif(\mathcal{A})} G_a\big) = \sum_{a \in \mathcal{A}} \frac{1}{|\mathcal{A}|}\overline{H}(G_{a})$. Thus, the function $P_A \mapsto \overline{H}\big(\bigsqcup_{a \in \mathcal{A}}^{P_A} G_a\big)$ is convex by Lemma \ref{th:etaconvex}, and satisfies \eqref{eq:lemma1A} with the interior point $P_A = \Unif(\mathcal{A})$: by Lemma \ref{lemma:convexanalysis} we have
\begin{align}
    \forall P_A \in \Delta(\mathcal{A}),\; \overline{H}\big(\textstyle\bigsqcup_{a \in \mathcal{A}}^{P_A} G_a\big) = \textstyle\sum_{a \in \mathcal{A}} P_A(a) \overline{H}(G_a).\label{eq:proofcormainB}
\end{align}

$(\Longleftarrow)$ Conversely, assume \eqref{eq:proofcormainB}, then $P_A \mapsto \overline{H}\big(\bigsqcup_{a \in \mathcal{A}}^{P_A} G_a\big)$ is linear. We can use Corollary \ref{cor:unif}, and we have $\overline{H}\big( \bigwedge_{a \in \mathcal{A}} G_a \big) = |\mathcal{A}|\overline{H}\big(\bigsqcup_{a \in \mathcal{A}}^{\Unif(\mathcal{A})} G_a\big) = \sum_{a \in \mathcal{A}} \overline{H}(G_{a})$.

\section{Proof of Theorem \ref{th:mainperfect}}\label{section:proofthmainperfect}

\subsection{Preliminary results}

Lemma \ref{lemma:Hkappasplit} comes from \cite[Corollary 3.4]{simonyi1995graph}, and states that the function $P_A \mapsto H_\kappa\big(\bigsqcup_{a \in \mathcal{A}}^{P_A} G_a \big)$, defined analogously to $P_A \mapsto \overline{H}\big(\bigsqcup_{a \in \mathcal{A}}^{P_A} G_a \big)$, is always linear. 
We give a proof of Lemma \ref{lemma:Hkappasplit} in Appendix \ref{section:prooflemmaHkappasplit} for the sake of completeness. The proof of Lemma \ref{lemma:perfectunion} is given in Appendix \ref{section:prooflemmaperfectunion}.

\begin{lemma}\label{lemma:Hkappasplit}
For all probabilistic graphs $(G_a)_{a \in \mathcal{A}}$ and $P_A \in \Delta(\mathcal{A})$, we have $H_\kappa\big(\bigsqcup_{a\in\mathcal{A}}^{P_A} G_a \big) = \sum_{a \in \mathcal{A}} P_A(a) H_\kappa(G_a)$.
\end{lemma}

\begin{lemma}\label{lemma:perfectunion}
    The probabilistic graph $\bigsqcup_{a\in\mathcal{A}}^{P_A} G_a$ is perfect if and only if  $G_a$ is perfect for all $a \in \mathcal{A}$.
\end{lemma}

\subsection{Main proof of Theorem \ref{th:mainperfect}}

For all $a \in \mathcal{A}$, let $G_a = (\mathcal{V}_a, \mathcal{E}_a, P_{V_a})$ be a perfect probabilistic graph. By Lemma \ref{lemma:perfectunion}, $\bigsqcup_{a \in \mathcal{A}}^{P_A} G_a$ is also perfect; and we have $\overline{H}\big(\bigsqcup_{a \in \mathcal{A}}^{P_A} G_a\big) = H_\kappa\big(\bigsqcup_{a \in \mathcal{A}}^{P_A} G_a\big)$ by Theorem \ref{th:Hbarperfect}. We also have $H_\kappa\big(\bigsqcup_{a \in \mathcal{A}}^{P_A} G_a\big) = \sum_{a \in \mathcal{A}} P_A(a) H_\kappa(G_a) = \sum_{a \in \mathcal{A}} P_A(a) \overline{H}(G_a)$ by Lemma \ref{lemma:Hkappasplit} and Theorem \ref{th:Hbarperfect} used on the perfect graphs $(G_a)_{a \in \mathcal{A}}$.

Therefore \eqref{eq:cormain01a} is satisfied by the graphs $(G_a)_{a \in \mathcal{A}}$ and $P_A$: by Theorem \ref{th:caraclin}, it follows that $\overline{H}\left(\bigwedge_{a \in \mathcal{A}} G_a \right) = \sum_{a \in \mathcal{A}} \overline{H}(G_a) = \sum_{a \in \mathcal{A}} H_\kappa(G_a)$, where the last equality comes from Theorem \ref{th:Hbarperfect}.

\section{Conclusion}

Theorem \ref{th:partvalues} shows that $\overline{H}\big(\bigsqcup_{a \in \mathcal{A}}^{P_A} G_a\big) = \frac{1}{k}\overline{H}\big(\bigwedge_{a \in \mathcal{A}} G_a^{\wedge k P_A(a)}\big)$ holds for all $P_A \in \Delta_k(\mathcal{A})$. The consequences of this result are stated in Theorem \ref{th:mainperfect}, Theorem \ref{th:caraclin} and Corollary \ref{cor:C_5}. We provide a single-letter formula for $\overline{H}$ for a new class of graphs. By \eqref{eq:debitopti}, this allows to characterize optimal rates for the two source coding problems depicted in Figure \ref{fig:Setting}.

\begin{proposition}\label{prop:optimalrate}
    The optimal rates in the settings from Figure \ref{fig:Setting}.a and Figure \ref{fig:Setting}.b are respectively given by $\overline{H}\big(\bigsqcup_{z \in \mathcal{Z}}^{P_{g(Y)}} G_{z}\big)$ and $\overline{H}\big(\bigwedge_{z \in \mathcal{Z}} G_{z}\big)$.
\end{proposition}

\appendices

\section{Proof of Corollary \ref{cor:C_5}}\label{section:proofcorC_5}

By \cite[Lemma 3]{tuncel2009complementary}, if $G$ is perfect then 
\begin{align}
    \overline{H}(G \overset{(\alpha, 1-\alpha)}{\sqcup} G_5) = \alpha \overline{H}(G) + (1-\alpha)\overline{H}(G_5).
\end{align}
By Theorem \ref{th:caraclin}, we have $\overline{H}(G \wedge G_5) = \overline{H}(G) + \overline{H}(G_5) = H_\kappa(G) + \frac{\log 5}{2}$; where the last equality comes from \cite[Example 1]{koulgi2003zero} which states that $\overline{H}(G_5) = \frac{\log 5}{2}$, and from Theorem \ref{th:mainperfect}.

\section{Proof dependencies}

\begin{figure}[h!]
    \centering
    \begin{tikzpicture}[scale=.9]
        \node[shape=rectangle, draw=black, fill=white] (DIST) at ($(2.5,0)$) {Lemma \ref{lemma:distrib}};
        
        \node[shape=rectangle, draw=black, fill=white] (INDUCED) at ($(0,0)$) {Lemma \ref{lemma:induced}};
        
        \node[shape=rectangle, draw=black, fill=white] (SIMPINDUCED) at ($(0,1)$) {Lemma \ref{lemma:simplificationinduced}};
        
        \node[shape=rectangle, draw=black, dashed, fill=white] (UNIONISO) at ($(5,0)$) {Lemma \ref{lemma:unionisomorph}};
        
        \node[shape=rectangle, draw=black, fill=white] (ETACONV) at ($(5,2)$) {Lemma \ref{th:etaconvex}};
        
        \node[shape=rectangle, draw=black, fill=white] (PARTTYPE) at ($(2.5,2)$) {Theorem \ref{th:partvalues}};
        
        \node[shape=rectangle, draw=black, fill=white] (PARTGEN) at ($(2.5,1)$) {Lemma \ref{lemma:partvalues}};
        
        \node[shape=rectangle, draw=black, fill=white] (TYPESPLIT) at ($(5,1)$) {Lemma \ref{lemma:typesplit}};
        
        \node[shape=rectangle, draw=black, dashed, fill=white] (HKAPPASPLIT) at ($(0,3)$) {Lemma \ref{lemma:Hkappasplit}};
        
        \node[shape=rectangle, draw=black, fill=white] (CONVEXAN) at ($(0,2)$) {Lemma \ref{lemma:convexanalysis}};
        
        \node[shape=rectangle, draw=black, fill=white] (CARACLIN) at ($(2.5,3)$) {Theorem \ref{th:caraclin}};
        
        \node[shape=rectangle, draw=black, fill=white] (MAINPERF) at ($(2.5,4)$) {Theorem \ref{th:mainperfect}};
        
        \node[shape=rectangle, draw=black, dashed, fill=white] (HBARPERF) at ($(-1,4)$) {Theorem \ref{th:Hbarperfect}};
        
        \node[shape=rectangle, draw=black, fill=white] (PERFUNION) at ($(5,3)$) {Lemma \ref{lemma:perfectunion}};
        
        \draw[->, >=stealth] (DIST) edge (PARTGEN);
        \draw[->, >=stealth] (UNIONISO) edge (PARTGEN);
        \draw[->, >=stealth] (INDUCED) edge (SIMPINDUCED);
        \draw[->, >=stealth] (SIMPINDUCED) edge (PARTGEN);
        \draw[->, >=stealth] (TYPESPLIT) edge (ETACONV);
        \draw[->, >=stealth] (PARTGEN) edge (ETACONV);
        \draw[->, >=stealth] (PARTGEN) edge (PARTTYPE);
        \draw[->, >=stealth] (PARTTYPE) edge (CARACLIN);
        \draw[->, >=stealth] (ETACONV) edge (CARACLIN);
        \draw[->, >=stealth] (CONVEXAN) edge (CARACLIN);
        \draw[->, >=stealth] (HKAPPASPLIT) edge (MAINPERF);
        \draw[->, >=stealth] (HBARPERF) edge (MAINPERF);
        \draw[->, >=stealth] (CARACLIN) edge (MAINPERF);
        \draw[->, >=stealth] (PERFUNION) edge (MAINPERF);
    \end{tikzpicture}
    \caption{An arrow from A to B means that A is used in the proof of B. Results from the literature are represented with a dashed outline.}
    \label{fig:dependances}
\end{figure}
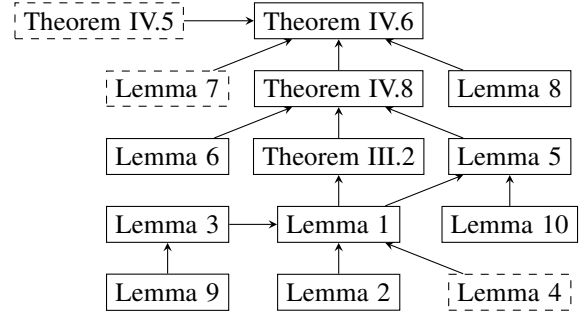

An illustration of the dependencies between the results can be found in Figure \ref{fig:dependances}.

\section{Proof of Lemma \ref{lemma:distrib}} \label{section:prooflemmadistrib}

    The probabilistic graphs in both sides of \eqref{eq:distribunion} have 
    \begin{align}
        \left(\bigsqcup_{a \in \mathcal{A}} \mathcal{V}_a\right) \times \left(\bigsqcup_{b \in \mathcal{B}} \mathcal{V}_b\right) = \bigsqcup_{(a,b) \in \mathcal{A}\times \mathcal{B}} \mathcal{V}_a \times \mathcal{V}_b
    \end{align}
    as set of vertices, with underlying distribution 
    \begin{align}
        & \left(\sum_{a \in \mathcal{A}} P_A(a) P_{V_a}\right)\left(\sum_{b \in \mathcal{B}} P_B(b) P_{V_b}\right) \nonumber\\
        & = \sum_{(a,b) \in \mathcal{A}\times \mathcal{B}} P_A(a) P_B(b) P_{V_a} P_{V_b}.
    \end{align}
    
    Now let us show that these two graphs have the same edges. Let $(v_{\mathcal{A}},v_{\mathcal{B}}),(v'_{\mathcal{A}},v'_{\mathcal{B}}) \in \left(\bigsqcup_{a \in \mathcal{A}} \mathcal{V}_a\right) \times \left(\bigsqcup_{b \in \mathcal{B}} \mathcal{V}_b\right)$; let $a_*, a'_* \in \mathcal{A}$ and $b_*,b'_* \in \mathcal{B}$ be the unique indexes such that
    \begin{align}
        (v_{\mathcal{A}},v_{\mathcal{B}}) \in \mathcal{V}_{a_*} \times \mathcal{V}_{b_*} \quad \text{and} \quad (v'_{\mathcal{A}},v'_{\mathcal{B}}) \in \mathcal{V}_{a'_*} \times \mathcal{V}_{b'_*}.
    \end{align}
    We have:
    \begin{align}
        &  \!\!\! (v_{\mathcal{A}},v_{\mathcal{B}}),(v'_{\mathcal{A}},v'_{\mathcal{B}}) \text{ are adjacent in } \left(\bigsqcup_{a \in \mathcal{A}}^{P_A} G_a\right) \wedge \left(\bigsqcup_{b \in \mathcal{B}}^{P_B} G_b\right)\\
        \Longleftrightarrow & \; v_{\mathcal{A}}, v'_{\mathcal{A}} \text{ adjacent in } \bigsqcup_{a \in \mathcal{A}}^{P_A} G_a \text{ and } \nonumber\\
        & v_{\mathcal{B}}, v'_{\mathcal{B}} \text{ adjacent in } \bigsqcup_{b \in \mathcal{B}}^{P_B} G_b\\
        \Longleftrightarrow & \; a_* = a'_* \text{ and } v_{\mathcal{A}} v'_{\mathcal{A}} \in \mathcal{E}_{a_*} \text{ and } b_* = b'_* \text{ and } v_{\mathcal{B}} v'_{\mathcal{B}} \in \mathcal{E}_{b_*} \\
        \Longleftrightarrow & \; (a_*,b_*) = (a'_*, b'_*) \text{ and } \nonumber\\ 
        & \; (v_{\mathcal{A}},v_{\mathcal{B}}),(v'_{\mathcal{A}},v'_{\mathcal{B}}) \text{ are adjacent in } G_{a_*} \wedge G_{b_*} \\
        \Longleftrightarrow & \; (v_{\mathcal{A}},v_{\mathcal{B}}),(v'_{\mathcal{A}},v'_{\mathcal{B}}) \text{ are adjacent in } \bigsqcup_{(a,b) \in \mathcal{A} \times \mathcal{B}}^{P_AP_B} G_a \wedge G_b.
    \end{align}

\section{Proof of Lemma \ref{lemma:simplificationinduced}} \label{section:prooflemmasimplificationinduced}

\subsection{Preliminary results}

In Lemma \ref{lemma:induced} we give upper and lower bounds on the chromatic entropy of an induced subgraph $G[\mathcal{S}]$, using the chromatic entropy of the whole graph $G$ and the probability $P_V(\mathcal{S})$. The core idea is that if $P_V(\mathcal{S})$ is close to $1$ and $H_\chi(G)$ is big, then $H_\chi(G[\mathcal{S}])$ is close to $H_\chi(G)$. The proof of Lemma \ref{lemma:induced} is given in Appendix \ref{section:prooflemmainduced}

\begin{lemma}\label{lemma:induced}
        Let $G = (\mathcal{V}, \mathcal{E}, P_V)$ and $\mathcal{S} \subset \mathcal{V}$, then 
        \begin{align}
            H_\chi(G) - 1 - (1-P_V(\mathcal{S}))\log|\mathcal{V}| \leq H_\chi(G[\mathcal{S}]) \leq\frac{H_\chi(G)}{P_V(\mathcal{S})}.
        \end{align}
\end{lemma}

\begin{remark}
    $H_\chi(G[\mathcal{S}])$ can be greater than $H_\chi(G)$, even if $G[\mathcal{S}]$ has less vertices and inherits the structure of $G$. This stems from the normalized distribution $P_V/P_V(\mathcal{S})$ on the vertices of $G[\mathcal{S}]$ which gives more weight to the vertices in $\mathcal{S}$. For example, consider
    \begin{align}
        G = \big(N_5, \Unif(\lbrace 1, ..., 5\rbrace)\big) \overset{(1-\epsilon,\epsilon)}{\sqcup} \big(K_5, \Unif(\lbrace 1, ..., 5\rbrace)\big),\nonumber
    \end{align}
    with $\mathcal{S}$ being the vertices in the connected component $K_5$ in $G$. Then $H_\chi(G) = \epsilon \log 5$ and $H_\chi(G[\mathcal{S}]) = \log 5$.
\end{remark}

\subsection{Main proof of Lemma \ref{lemma:simplificationinduced}}

By Lemma \ref{lemma:induced}, we have for all $n \in \mathbb{N}^\star$:
\begin{align}
    & H_\chi(G^{\wedge n}) - 1 - (1-P^n_V(\mathcal{S}^n))\log|\mathcal{V}| \nonumber\\
    \leq \: & H_\chi(G^{\wedge n}[\mathcal{S}^n]) \leq \frac{H_\chi(G^{\wedge n})}{P^n_V(\mathcal{S}^n)}.
\end{align}
Since $P^n_V(\mathcal{S}^n) \rightarrow 1$, and $H_\chi(G^{\wedge n}) = n \overline{H}(G) + o(n)$ when $n \rightarrow \infty$, the desired results follows immediately by normalization and limit.

\section{Proof of Lemma \ref{th:etaconvex}}\label{section:proofetaconvex}

\subsection{Preliminary results}

Lemma \ref{lemma:typesplit} is a generalization for infinite sequences of the following observation: if $T_{\overline{a}^n} = P_A \in \Delta_n(\mathcal{A})$ satisfies $P_A = \frac{i}{n}P'_A + \frac{n-i}{n}P''_A$ with $P'_A \in \Delta_i(\mathcal{A})$ and $P''_A \in \Delta_{n-i}(\mathcal{A})$, then $\overline{a}^n$ can be separated into two subsequences $a'^i$ and $a''^{n-i}$ such that $T_{a'^i} = P'_A$ and $T_{a''^{n-i}} = P''_A$. The proof is given in Appendix \ref{section:prooflemmatypesplit}.

\begin{lemma}[Type-splitting lemma]\label{lemma:typesplit}
    Let $(\overline{a}_n)_{n \in \mathbb{N}^\star} \in \mathcal{A}^{\mathbb{N}^\star}$ be a sequence such that $T_{\overline{a}^n} \rightarrow P_A \in \Delta(\mathcal{A})$ when $n \rightarrow \infty$, let $\beta \in (0,1)$ and $P'_A, P''_A \in \Delta(\mathcal{A})$ such that
    \begin{align}
        P_A = \beta P'_A + (1-\beta)P''_A.
    \end{align}
    Then there exists a sequence $(b_n)_{n \in \mathbb{N}^\star} \in \lbrace 0, 1 \rbrace^{\mathbb{N}^\star}$ such that the two extracted sequences $a' \doteq (\overline{a}_n)_{\substack{n\in \mathbb{N}^\star, \\ b_n = 0}}$ and $a'' \doteq (\overline{a}_n)_{\substack{n\in \mathbb{N}^\star, \\ b_n = 1}}$ satisfy
    \begin{align}
        & T_{b^n} \underset{n \rightarrow \infty}{\rightarrow} (\beta, 1-\beta), & & \\
        & T_{a'^n} \underset{n \rightarrow \infty}{\rightarrow} P'_A, & & T_{a''^n} \underset{n \rightarrow \infty}{\rightarrow} P''_A.
    \end{align}
\end{lemma}

\subsection{Main proof of Lemma \ref{th:etaconvex}}

\textbf{($\eta$ Lipschitz)} Let us first prove that $\eta$ is Lipschitz. For all $P_A, P'_A \in \Delta(\mathcal{A})$ we need to bound the quantity $|\eta(P_A)-\eta(P'_A)|$; by Lemma \ref{lemma:partvalues} this is equivalent to bounding
\begin{align}
     \lim_{n \rightarrow \infty} \frac{1}{n}\left|H_\chi\left(\bigwedge_{a \in \mathcal{A}} G_a^{\wedge n T_{\overline{a}^n} (a)}\right) - H_\chi\left(\bigwedge_{a \in \mathcal{A}} G_a^{\wedge n T_{\overline{a}'^n} (a)}\right)\right|\label{eq:prooflemmaetaconvexA0}
    \end{align}
where $(T_{\overline{a}^n}, T_{\overline{a}'^n}) \rightarrow (P_A, P'_A)$ when $n \rightarrow \infty$.

Fix $n \in \mathbb{N}^\star$, we assume that the quantity inside $|\cdot|$ in \eqref{eq:prooflemmaetaconvexA0} is positive; the other case can be treated with the same arguments by symmetry of the roles. We have
\begin{align}
    & H_\chi\left(\bigwedge_{a \in \mathcal{A}} G_a^{\wedge n T_{\overline{a}^n} (a)}\right) - H_\chi\left(\bigwedge_{a \in \mathcal{A}} G_a^{\wedge n T_{\overline{a}'^n} (a)}\right) \label{eq:Th3A}\\
    \leq & \; H_\chi\left(\bigwedge_{a \in \mathcal{A}} G_a^{\wedge n T_{\overline{a}^n} (a)}\right) - H_\chi\Bigg(\bigwedge_{a \in \mathcal{A}} G_a^{\wedge n \min(T_{\overline{a}^n}(a),T_{\overline{a}'^n}(a))}\! \Bigg) \label{eq:Th3B}\\
    = & \; H_\chi\Bigg(\!\bigwedge_{a \in \mathcal{A}} G_a^{\wedge n \min(T_{\overline{a}^n}(a),T_{\overline{a}'^n}(a))} \bigwedge_{a \in \mathcal{A}} G_a^{\wedge n |T_{\overline{a}^n}(a)-T_{\overline{a}'^n}(a)|_+}\!\!\Bigg) \nonumber\\
    & - H_\chi\Bigg(\bigwedge_{a \in \mathcal{A}} G_a^{\wedge n \min(T_{\overline{a}^n}(a),T_{\overline{a}'^n}(a))} \Bigg) \label{eq:Th3C}\\
    \leq & \; H_\chi\Bigg(\bigwedge_{a \in \mathcal{A}} G_a^{\wedge n |T_{\overline{a}^n}(a)-T_{\overline{a}'^n}(a)|_+}\Bigg) \label{eq:Th3D}\\
    \leq & \; \log\left(\max_a |\mathcal{V}_a|\right) \sum_{a \in \mathcal{A}} n |T_{\overline{a}^n}(a)-T_{\overline{a}'^n}(a)|_+ \label{eq:Th3E}\\
    \leq & \; n \log\left(\max_a |\mathcal{V}_a|\right) \|T_{\overline{a}^n}-T_{\overline{a}'^n}\|_1, \label{eq:Th3F}
\end{align}
where $|\cdot|_+ = \max(\cdot,0)$ and $\|T_{\overline{a}^n}-T_{\overline{a}'^n}\|_1 = \sum_{a \in \mathcal{A}} |T_{\overline{a}^n}(a)-T_{\overline{a}'^n}(a)|$; \eqref{eq:Th3B} follows from the removal of terms in the second product, as $H_\chi(G\wedge G') \geq H_\chi(G)$ for all probabilistic graphs $G,G'$; \eqref{eq:Th3C} is an arrangement of the terms in the first product, as $\min(s,t) + \max(s-t,0) = s$ for all real numbers $s,t$; \eqref{eq:Th3D} comes from the subadditivity of $H_\chi$; \eqref{eq:Th3E} follows from $H_\chi(G_a) \leq \log \max_{a'} |\mathcal{V}_{a'}|$ for all $a \in \mathcal{A}$; \eqref{eq:Th3F} results from $|T_{\overline{a}^n}(a)-T_{\overline{a}'^n}(a)|_+ \leq |T_{\overline{a}^n}(a)-T_{\overline{a}'^n}(a)|$ for all $a\in \mathcal{A}$.

By normalization and limit, it follows that
\begin{align}
    |\eta(P_A)-\eta(P'_A)| & \leq \lim_{n \rightarrow \infty} \log\left(\max_a |\mathcal{V}_a|\right)\cdot\|T_{\overline{a}^n}-T_{\overline{a}'^n}\|_1 \\
    & = \log\left(\max_a |\mathcal{V}_a|\right)\cdot\|P_A - P'_A\|_1.\label{eq:Th3G}
\end{align}
Hence $\eta$ is $(\log \max_a |\mathcal{V}_a|)$-Lipschitz.\newline

\textbf{($\eta$ convex)} Let us now prove that $\eta$ is convex. Let $P'_A, P''_A \in \Delta(\mathcal{A})$, and $\beta \in (0,1)$, we have by Lemma \ref{lemma:partvalues}
\begin{align}
    \eta\big(\beta P'_A + (1-\beta) P''_A\big) = \lim_{n \rightarrow \infty} \frac{1}{n} H_\chi\left(\bigwedge_{a \in \mathcal{A}} G_a^{\wedge n T_{\overline{a}^n} (a)}\right),\label{eq:Th3H}
\end{align}

where $T_{\overline{a}^n} \rightarrow \beta P'_A + (1-\beta) P''_A$ when $n \rightarrow \infty$. By Lemma \ref{lemma:typesplit}, there exists $(b_n)_{n \in \mathbb{N}^\star} \in \lbrace 0,1\rbrace^{\mathbb{N}^\star}$ such that the decomposition of $(\overline{a}_n)_{n \in \mathbb{N}^\star}$ into two subsequences $a' \doteq (\overline{a}_n)_{\substack{n\in \mathbb{N}^\star, \\ b_n = 0}}$ and $a'' \doteq (\overline{a}_n)_{\substack{n\in \mathbb{N}^\star, \\ b_n = 1}}$ satisfies
    \begin{align}
        & T_{b^n} \underset{n \rightarrow \infty}{\rightarrow} (\beta,1-\beta), & &  \label{eq:Th3J}\\
        & T_{a'^n} \underset{n \rightarrow \infty}{\rightarrow} P'_A, & & T_{a''^n} \underset{n \rightarrow \infty}{\rightarrow} P''_A.\label{eq:Th3K}
    \end{align}

For all $n \in \mathbb{N}^\star$, let $\Xi(n) \doteq n T_{b^n}(0)$, we have
\begin{align}
    & \eta\big(\beta P'_A + (1-\beta) P''_A\big) \\
    = & \; \lim_{n \rightarrow \infty} \frac{1}{n} H_\chi\left(\bigwedge_{a \in \mathcal{A}} G_a^{\wedge \Xi(n) T_{a'^{\Xi(n)}}(a) + (n-\Xi(n)) T_{a''^{n-\Xi(n)}}(a) } \right)\label{eq:Th3L} \\
    \leq & \; \lim_{n \rightarrow \infty} \frac{\Xi(n)}{n} \frac{1}{\Xi(n)} H_\chi\left(\bigwedge_{a \in \mathcal{A}} G_a^{\wedge \Xi(n) T_{a'^{\Xi(n)}}(a)}\right) \\
    & + \frac{n-\Xi(n)}{n} \frac{1}{n-\Xi(n)}  H_\chi\left(\bigwedge_{a \in \mathcal{A}} G_a^{\wedge (n-\Xi(n)) T_{a''^{n-\Xi(n)}}(a)}\right) \label{eq:Th3M}\\
    = & \; \beta \eta(P'_A) + (1-\beta)\eta(P''_A); \label{eq:Th3O}
\end{align}
where \eqref{eq:Th3L} comes from \eqref{eq:Th3H}; \eqref{eq:Th3M} follows from the subadditivity of $H_\chi$; \eqref{eq:Th3O} comes from \eqref{eq:Th3J}, \eqref{eq:Th3K} and Lemma \ref{lemma:partvalues}. Since \eqref{eq:Th3O} holds for all $P'_A, P''_A \in \Delta(\mathcal{A})$ and $\beta \in (0,1)$, we have that $\eta$ is convex.

\section{Proof of Lemma \ref{lemma:convexanalysis}} \label{section:prooflemmaconvexanalysis}
It can be easily observed that
\begin{align}
    & \exists P_A \in \interior(\Delta(\mathcal{A})), \, \gamma(P_A) = \sum_{a \in \mathcal{A}} P_A(a) \gamma(\mathds{1}_a) \label{eq:prooflemmaconvexanA1}\\
    \Longleftarrow \;\; & \forall P_A \in \Delta(\mathcal{A}), \, \gamma(P_A) = \sum_{a \in \mathcal{A}} P_A(a) \gamma(\mathds{1}_a). \label{eq:prooflemmaconvexanA2}
\end{align}
Now let us prove \eqref{eq:prooflemmaconvexanA1} $\Rightarrow$ \eqref{eq:prooflemmaconvexanA2}. Let $P^*_A \in \interior\Delta(\mathcal{A})$ such that $\gamma(P^*_A) = \sum_{a \in \mathcal{A}} P^*_A(a) \gamma(\mathds{1}_a)$. Let $m : \Delta(\mathcal{A}) \rightarrow \mathbb{R}$ linear such that $m(P^*_A) = \gamma(P^*_A)$ and $\forall P_A \in \Delta(\mathcal{A}),\, m(P_A) \leq \gamma(P_A)$. We have
\begin{align}
    0 = \gamma(P^*_A) - m(P^*_A) = \sum_{a \in \mathcal{A}} P^*_A(a) \big(\gamma(\mathds{1}_a) - m(\mathds{1}_a)\big);
\end{align}
and therefore $\gamma(\mathds{1}_a) = m(\mathds{1}_a)$ for all $a \in \mathcal{A}$, as $\gamma - m \geq 0$ and $P^*_A(a) > 0$ for all $a \in \mathcal{A}$. For all $P_A \in \Delta(\mathcal{A})$, we have
\begin{align}
    f(P_A) & \leq \sum_{a \in \mathcal{A}} P_A(a) \gamma(\mathds{1}_a) \\
    & = \sum_{a \in \mathcal{A}} P_A(a) m(\mathds{1}_a) = m(P_A),
\end{align}
hence $\gamma = m$ and $\gamma$ is linear.

\section{Proof of Lemma \ref{lemma:Hkappasplit}}\label{section:prooflemmaHkappasplit}
Let $G_a = (\mathcal{V}_a, \mathcal{E}_a, P_{V_a})$, and $G = (\mathcal{V}, \mathcal{E},P_{V})$ such that $G = \bigsqcup_{a \in \mathcal{A}}^{P_A} G_a$. Let $A$ be the random variable with distribution $P_A$ such that $V = V_A$, i.e. $P_{V|A = a} = P_{V_a}$.\newline

\textit{Achievability}

For all $a \in \mathcal{A}$, let $W^*_a$ be a minimizer of
\begin{align}
    \min_{V_a \in W_a \in \Gamma(G_a)} I(V_a;W_a).\label{eq:defWastar}
\end{align}
Let $W^*$ be the random variable defined as follows: for all $\mathcal{S} \in \Gamma(G)$, $a \in \mathcal{A}$ and $v_a \in \mathcal{V}_a$,
\begin{align}
    P_{W^*| A = a, V = v_a}(\mathcal{S}) \doteq P_{W^*_a|V_a = v_a}(\mathcal{S}_a) \prod_{\substack{a' \in \mathcal{A} \\ a' \neq a}} P_{W^*_{a'}}(\mathcal{S}_{a'}), \label{eq:defWstar}
\end{align}
where $\mathcal{S}$ is uniquely decomposed as $\bigsqcup_{a \in \mathcal{A}} \mathcal{S}_a$, with $\mathcal{S}_a \in \Gamma(G_a)$ for all $a \in \mathcal{A}$. The random variable $W^*$ takes its values in $\Gamma(G)$, as
\begin{align}
    P_{W^*}(\mathcal{S}) > 0 \Longrightarrow \forall a \in \mathcal{A}, \, P_{W^*_a}(\mathcal{S}_a) > 0.
\end{align}
The conditional distribution w.r.t. $(A = a)$ writes:
\begin{align}
    P_{W^*| A = a}(\mathcal{S}) = & \sum_{v_a \in \mathcal{V}_a} P_{V_a}(v_a) P_{W^*| A = a, V = v_a}(\mathcal{S}) \label{eq:proofHkappasplitW2}\\
    = & \sum_{v_a \in \mathcal{V}_a} P_{V_a}(v_a) P_{W^*_a|V_a = v_a}(\mathcal{S}_a) \prod_{\substack{a' \in \mathcal{A} \\ a' \neq a}} P_{W^*_{a'}}(\mathcal{S}_{a'}) \label{eq:proofHkappasplitW3}\\
    = & \prod_{a' \in \mathcal{A}} P_{W^*_{a'}}(\mathcal{S}_{a'}). \label{eq:proofHkappasplitW4}
\end{align}
It follows that the random variable $W^*$ is independent of $A$ as the expression \eqref{eq:proofHkappasplitW4} does not depend on $a$. Note that $P_{W^*}$ defined in \eqref{eq:defWstar} is a probability distribution, as 
\begin{align}
    \sum_{\mathcal{S} \in \Gamma(G)} P_{W^*}(\mathcal{S}) = & \; \sum_{\mathcal{S} \in \Gamma(G)} \;\;\prod_{a' \in \mathcal{A}} P_{W^*_{a'}}(\mathcal{S}_{a'}) \label{eq:proofHkappasplitX2}\\
    = & \; \sum_{(\mathcal{S}_{a'})_{a' \in \mathcal{A}} \in \prod_{a' \in \mathcal{A}} \Gamma(G_{a'})} \;\; \prod_{a' \in \mathcal{A}}P_{W^*_{a'}}(\mathcal{S}_{a'}) \label{eq:proofHkappasplitX3}\\
    = & \; 1, \label{eq:proofHkappasplitX4}
\end{align}
where \eqref{eq:proofHkappasplitX2} comes from \eqref{eq:proofHkappasplitW4}; \eqref{eq:proofHkappasplitX3} follows from $\Gamma(G) = \big\lbrace \bigsqcup_{a \in \mathcal{A}} \mathcal{S}_a \:\big|\: \forall a \in \mathcal{A}, \, \mathcal{S}_a \in \Gamma(G_a) \big\rbrace$; and \eqref{eq:proofHkappasplitX4} holds as $W^*_a$ takes its values in $\Gamma(G_a)$ for all $a \in \mathcal{A}$.

Now, let us show that $V \in W^*$ with probability one. For all $a \in \mathcal{A}$ and $v_a \in \mathcal{V}_a$,
\begin{align}
    \lbrace \mathcal{S} \cap \mathcal{V}_a \:|\: \mathcal{S} \in \supp P_{W^*|V = v_a} \rbrace = \supp P_{W^*_a|V_a = v_a};
\end{align}
where $\supp$ denotes the support of a probability distribution. Since $V_a \in W^*_a$ with probability one, all the sets in $\supp P_{W^*_a|V_a = v_a}$ contain $v_a$, hence all sets in $\supp P_{W^*|V = v_a}$ also contain $v_a$: $V \in W^*$ with probability one.

Now let us combine the results on $W^*$:
\begin{align}
    H_\kappa(G) & \leq I(V;W^*) \label{eq:proofHkappasplitZ1}\\
    & = I(V,A;W^*) \label{eq:proofHkappasplitZ2}\\
    & = I(A;W^*) + \sum_{a \in \mathcal{A}} P_A(a) I(V;W^*|A = a)\label{eq:proofHkappasplitZ3}\\
    & = \sum_{a \in \mathcal{A}} P_A(a) I(V_a;W_a^*) \label{eq:proofHkappasplitZ4}\\
    & = \sum_{a \in \mathcal{A}} P_A(a) H_\kappa(G_a); \label{eq:proofHkappasplitZ5}
\end{align}
where \eqref{eq:proofHkappasplitZ1} holds as $W^*$ takes its values in $\Gamma(G)$ and $V \in W^*$ with probability one; \eqref{eq:proofHkappasplitZ2} holds as $A$ is a deterministic function of $V$; \eqref{eq:proofHkappasplitZ3} comes from the decomposition $I(V,A;W) = I(V;W|A) + I(A;W)$; \eqref{eq:proofHkappasplitZ4} follows from the independence of $A$ and $W^*$; \eqref{eq:proofHkappasplitZ5} comes from the fact that $W^*_a$ minimizes \eqref{eq:defWastar}. \newline

\textit{Converse}
\begin{align}
    \!\!H_\kappa\left(\bigsqcup_{a \in \mathcal{A}}^{P_A} G_a \right) &  = \min_{V \in W \in \Gamma(G)} I(V,A;W) \label{eq:proofHkappasplitA2}\\
    & \geq \min_{V \in W \in \Gamma(G)} \sum_{a \in \mathcal{A}} P_A(a) I(V;W|A = a) \label{eq:proofHkappasplitA3}\\
    & = \sum_{a \in \mathcal{A}} P_A(a) \min_{V \in W \in \Gamma(G)} \! I(V;W|A = a) \label{eq:proofHkappasplitA4}\\
    & = \sum_{a \in \mathcal{A}} P_A(a) \min_{V_a \in W \in \Gamma(G_a)} I(V_a;W) \label{eq:proofHkappasplitA5}\\
    & = \sum_{a \in \mathcal{A}} P_A(a) H_\kappa(G_a); \label{eq:proofHkappasplitA6}
\end{align}
where \eqref{eq:proofHkappasplitA2} holds as $A$ is a deterministic function of $V$; \eqref{eq:proofHkappasplitA3} follows from the decomposition $I(V,A;W) = I(V;W|A) + I(A;W)$; and \eqref{eq:proofHkappasplitA5} holds as $V = V_A$.

\section{Proof of Lemma \ref{lemma:perfectunion}}\label{section:prooflemmaperfectunion}
$(\Longrightarrow)$ Let $G = \bigsqcup_{a \in \mathcal{A}}^{P_A} G_a$ be a perfect probabilistic graph. Let $a' \in \mathcal{A}$ and $\mathcal{S}_{a'} \subset \mathcal{V}_{a'}$. We have $\chi\big(\big(\bigsqcup_{a \in \mathcal{A}}^{P_A} G_a \big)[\mathcal{S}_{a'}]\big) = \omega\big(\big(\bigsqcup_{a \in \mathcal{A}}^{P_A} G_a \big)[\mathcal{S}_{a'}]\big)$ since $G$ is perfect, and therefore $\chi(G_{a'}[\mathcal{S}_{a'}]) = \omega(G_{a'}[\mathcal{S}_{a'}])$, as $\big(\bigsqcup_{a \in \mathcal{A}}^{P_A} G_a \big)[\mathcal{S}_{a'}] = G_{a'}[\mathcal{S}_{a'}]$. Thus all the graphs $(G_a)_{a \in \mathcal{A}}$ are perfect.

$(\Longleftarrow)$ Conversely, assume that for all $a \in \mathcal{A}$, $G_a = (\mathcal{V}_a, \mathcal{E}_a, P_{V_a})$ is perfect. Then for all $\mathcal{S} \subset \bigsqcup_{a \in \mathcal{A}} \mathcal{V}_a$, $\mathcal{S}$ can be written as $\bigsqcup_{a \in \mathcal{A}} \mathcal{S}_a$ where $\mathcal{S}_a \subset \mathcal{V}_a$ for all $a \in \mathcal{A}$, and we have for all $P_A \in \Delta(\mathcal{A})$:
\begin{align}
    \chi\left(\left(\bigsqcup_{a \in \mathcal{A}}^{P_A} G_a \right)[\mathcal{S}]\right) & = \chi\left(\bigsqcup_{a \in \mathcal{A}}^{P_A} G_a[\mathcal{S}_a] \right) \label{eq:proofmainperfectZ1}\\
    & = \max_{a \in \mathcal{A}} \chi\left(G_a[\mathcal{S}_a]\right) \label{eq:proofmainperfectZ2}\\
    & = \max_{a \in \mathcal{A}} \omega\left(G_a[\mathcal{S}_a]\right), \label{eq:proofmainperfectZ3}
\end{align}
and similarly, $\omega\big(\big(\bigsqcup_{a \in \mathcal{A}}^{P_A} G_a\big)[\mathcal{S}]\big) = \max_{a \in \mathcal{A}} \omega\left(G_a[\mathcal{S}_a]\right)$.
Hence $\bigsqcup_{a \in \mathcal{A}}^{P_A} G_a$ is also perfect.

\section{Proof of Lemma \ref{lemma:induced}}\label{section:prooflemmainduced}

Let $c^* : \mathcal{V} \rightarrow \mathcal{C}$ and $c^*_\mathcal{S} : \mathcal{S} \rightarrow \mathcal{C}$ be the optimal colorings of $G$ and $G[\mathcal{S}]$, respectively. Consider the coloring $c : \mathcal{V} \rightarrow \mathcal{C} \sqcup \mathcal{V}$ of $G$ defined by $c(v) = c^*_\mathcal{S}$ if $v \in \mathcal{S}$, $c(v) = v$ otherwise.\newline

\textbf{(Lower bound)} On one hand, we have
\begin{align}
    H_\chi(G) \leq &\: H(c(V), \mathds{1}_{V \in \mathcal{S}}) \label{eq:prooflemmainducedA1}\\
    = &\: H(\mathds{1}_{V \in \mathcal{S}}) + P_V(\mathcal{S}) H(c(V)|V \in \mathcal{S}) \nonumber\\
     &+ (1-P_V(\mathcal{S})) H(c(V)|V \notin \mathcal{S}) \label{eq:prooflemmainducedA2}\\
    \leq &\: 1 + H(c^*_\mathcal{S}(V)|V \in \mathcal{S}) + (1-P_V(\mathcal{S})) \log |\mathcal{V}| \label{eq:prooflemmainducedA3}\\
    = &\: H_\chi(G[\mathcal{S}]) + 1 + (1-P_V(\mathcal{S})) \log |\mathcal{V}|;\label{eq:prooflemmainducedA4}
\end{align}
where \eqref{eq:prooflemmainducedA1} comes from the fact that $c$ is a coloring of $G$; \eqref{eq:prooflemmainducedA2} is a decomposition using conditional entropies; \eqref{eq:prooflemmainducedA3} comes from the construction of $c$: $c|_\mathcal{S} = c^*_{\mathcal{S}}$; \eqref{eq:prooflemmainducedA4} follows from the optimality of $c^*_\mathcal{S}$ as a coloring of $G[\mathcal{S}]$.\newline

\textbf{(Upper bound)} On the other hand, 
\begin{align}
    & H_\chi(G[\mathcal{S}]) \nonumber\\
    \leq & \: H(c^*(V) | V \in \mathcal{S}) \label{eq:prooflemmainducedB1} \\
    = & \, \frac{1}{P_V(\mathcal{S})}\Big(\!H(c^*(V)| \mathds{1}_{V \in \mathcal{S}}) - (1-P_V(\mathcal{S}))H(c^*(V) | V \notin \mathcal{S})\!\Big) \label{eq:prooflemmainducedB2} \\
    \leq & \: \frac{H(c^*(V))}{P_V(\mathcal{S})} = \frac{H_\chi(G)}{P_V(\mathcal{S})} \label{eq:prooflemmainducedB3} 
\end{align}
where \eqref{eq:prooflemmainducedB1} comes from the fact that $c^*$ induces a coloring of $G[\mathcal{S}]$; \eqref{eq:prooflemmainducedB2} is a decomposition using conditional entropies; \eqref{eq:prooflemmainducedB3} results from the elimination of negative terms and the optimality of $c^*$.

\section{Proof of Lemma \ref{lemma:typesplit}}\label{section:prooflemmatypesplit}

Let $(\overline{a}_n)_{n \in \mathbb{N}^\star} \in \mathcal{A}^{\mathbb{N}^\star}$ be a sequence such that $T_{\overline{a}^n} \rightarrow P_A = \beta P'_A + (1-\beta)P''_A$ when $n \rightarrow \infty$.

Consider a sequence $(B_n)_{n \in \mathbb{N}^\star}$ of independent Bernoulli random variables such that for all $n \in \mathbb{N}^\star$,
\begin{align}
    \Pr(B_n = 0) = \frac{\beta P'_A(\overline{a}_n)}{P_A(\overline{a}_n)}.
\end{align}
By the strong law of large numbers,
\begin{align}
    \Pr\left(T_{B^n,\overline{a}^n} \underset{n \rightarrow \infty}{\rightarrow} (\beta P'_A, (1-\beta)P''_A) \right) = 1.
\end{align}
Therefore, there exists at least one realization $(b_n)_{n \in \mathbb{N}^\star}$ of $(B_n)_{n \in \mathbb{N}^\star}$ such that $T_{b^n,\overline{a}^n}$ converges to $\big(\beta P'_A, (1-\beta)P''_A \big)$. The convergences of marginal and conditional types yield
    \begin{align}
        & T_{b^n} \underset{n \rightarrow \infty}{\rightarrow} (\beta, 1-\beta),  \\
        & T_{a'^n} \underset{n \rightarrow \infty}{\rightarrow} P'_A, \qquad T_{a''^n} \underset{n \rightarrow \infty}{\rightarrow} P''_A,
    \end{align}
where $a' \doteq (\overline{a}_n)_{\substack{n\in \mathbb{N}^\star, \\ b_n = 0}}$ and $a'' \doteq (\overline{a}_n)_{\substack{n\in \mathbb{N}^\star, \\ b_n = 1}}$ are the extracted sequences.

\end{document}